\documentclass{mem}
\usepackage{natbib}\usepackage{txfonts}\usepackage{balance}
\usepackage{graphicx}
\usepackage[a4paper,breaklinks,dvipdfm]{hyperref}
\idline{75}{282}
\begin{document}
\def\teff{$T\rm_{eff }$}
\def\kms{$\mathrm {km s}^{-1}$}

\title{
Terzan 5: a pristine fragment of the Galactic Bulge?
}

   \subtitle{}

\author{
D. Massari\inst{1},
F. R. Ferraro\inst{1},
A. Mucciarelli\inst{1},
L. Origlia\inst{2},
E. Dalessandro\inst{1},
B. Lanzoni\inst{1}
          }

  \offprints{D. Massari}

\institute{
Dipartimento di Fisica ed Astronomia --
Viale Berti-Pichat 6/2,
I-40127, Bologna, Italy\\ \email{davide.massari@unibo.it}
\and
INAF-Osservatorio Astronomico di Bologna, via
  Ranzani 1, 40127, Bologna, Italy
}

\authorrunning{Massari }

\titlerunning{Terzan 5: a pristine fragment of the Galactic Bulge?}

\abstract{
Terzan 5 is a stellar system located in the inner Bulge of the Galaxy and has been
historically catalogued as a globular cluster. However, recent photometric 
(\citealt{f09}) and spectroscopic (\citealt{origlia,o13}) investigations 
have shown that it hosts at least three stellar populations with different iron 
abundances (with a total spread of $\Delta$[Fe/H]$>1$ dex) thus demonstrating that 
Terzan 5 is not a genuine globular cluster. In addition, the striking similarity between 
the chemical patterns of this system and those of its surrounding environment, the Galactic Bulge,
from the point of view of both the metallicity distribution and the $\alpha-$element 
enrichment, suggests that Terzan 5 could be a pristine fragment of the Bulge itself.
}
\maketitle{}

\section{General framework}

Terzan 5 is a stellar system located in the inner Bulge of the Galaxy, at a distance
of 5.9 kpc \citep{valenti}. Since its discovery in 1962 it has been catalogued as
a globular cluster (GC). 
Terzan 5 resides in an extremely extincted region of the sky, with an average
color excess E(B-V)=2.38 mag (\citealt{valenti}) which varies spatially by more than 
0.7 mag in the cluster central regions (\citealt{massari}).

It hosts 34 millisecond pulsars (MSPs), which corresponds to more than                  
the 25\% of the entire population of MSPs known to date in Galactic GCs (see 
\citealt{ransom}). Terzan 5 has also the largest value of the collisional parameter
$\Gamma$ (\citealt{vhut}; \citealt{l10}) among all Galactic stellar systems.

\begin{figure}[!htp]
\resizebox{\hsize}{!}{\includegraphics[clip=true]{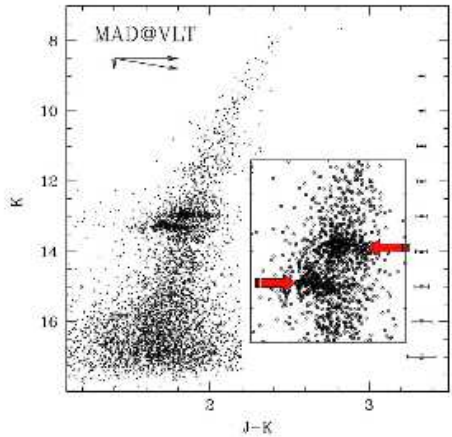}}
\caption{
\footnotesize
The two HBs of Terzan 5. In the main panel,
the (K, J-K) CMD of the central region of Terzan 5. In the inset,
a magnified view of the HB region, with the two RCs marked
with (red in the online version) arrows. Error bars are also plotted at different magnitude
levels. For details see \cite{f09}.
}
\label{hbs}
\end{figure}

\begin{figure}[!htp]
\resizebox{\hsize}{!}{\includegraphics[clip=true]{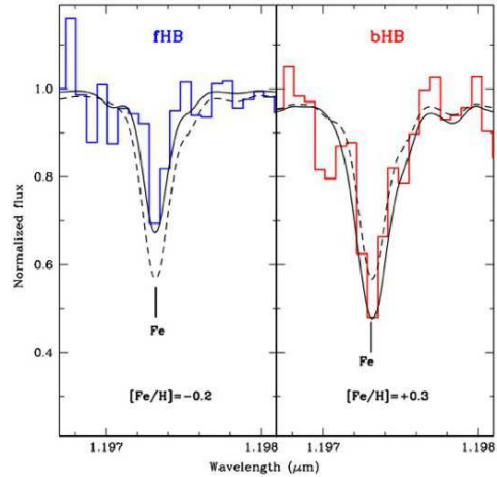}}
\caption{
\footnotesize
J-band spectra near the 1.1973 $\mu$m iron line for faint HB (left)
and bright HB (right) stars. The black solid lines correspond to the
best fit synthetic spectra obtained for temperatures and gravities
derived from evolutionary models reproducing the observed colours of 
the RCs stars (for details see \citealt{f09}).
}
\label{irons}
\end{figure}

Propelled by these peculiarities, a detailed investigation of this system has been performed by means of
adaptive optics observations with MAD@VLT by \cite{f09}. These authors discovered the presence of two 
clearly separated horizontal branches (HBs) in the infrared (IR) color magnitude diagram (CMD) of Terzan 5 (see Fig. 
\ref{hbs}). A prompt spectroscopic follow up performed with the near IR spectrograph 
NIRSPEC mounted at the Keck II telescope, demonstrated that both the populations are cluster members
(based on the measured radial velocities), and that they have 
different iron abundances, with $\Delta$[Fe/H]$\simeq0.5$ dex (Fig. \ref{irons}).
A detailed study on a larger sample of 33 giants performed by \cite{origlia} confirmed   
that the metal-poor population has an average [Fe/H]$=-0.25$ dex, while the metal-rich
population is enriched to [Fe/H]$=+0.27$ dex. 
In addition, the two populations have different radial distributions (\citealt{f09}),
with the metal-rich one being more centrally segregated.
This evidence, coupled with the large difference in the iron abundances
is a clear hint of self-enrichment. Therefore, Terzan 5 probably experienced complex 
formation and evolution, and its initial mass likely was much larger
than the current one ($10^6 M_{\odot}$, see \citealt{l10}) in order to
retain the iron-enriched gas ejected by supernova (SN) explosions.

\begin{figure}[!htp]
\resizebox{\hsize}{!}{\includegraphics[clip=true]{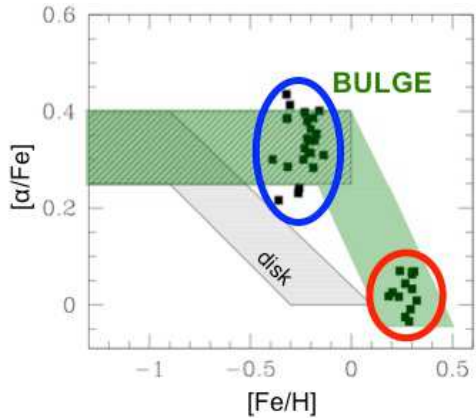}}
\caption{
\footnotesize
[$\alpha$/Fe] vs. [Fe/H] abundance ratios for Terzan 5 giants. The behavior
of the two populations follows that of the bulge stars (green region),
suggesting a strong evolutionary link between Terzan 5 and the Bulge itself.
}
\label{alfafe}
\end{figure}

A key ingredient to understand the true nature of Terzan 5 comes from the $\alpha$-elements.
Since they are efficiently produced and injected in the inter-stellar medium by type II SNe, 
they trace the enrichment history of a 
stellar system providing important clues on its star formation rate (SFR) and initial mass function
(IMF). The study by \cite{origlia} showed that while the
metal-poor component is $\alpha$-enhanced to values of $[\alpha$/Fe]$=+0.34$, 
the metal rich population has a solar scaled abundance ($[\alpha$/Fe]$=+0.03$ dex).
Such a behavior suggests that a very high SFR supported the formation and evolution
of Terzan 5. Moreover, as shown in Fig.\ref{alfafe}, such a behavior 
is completely different from the typical behavior observed for Halo or Disk stars, 
but it is strikingly similar to the behavior observed for Terzan 5 environment, the Bulge.

These observational results clearly demonstrate that Terzan 5 is not a genuine GC, but a complex
stellar system that formed and evolved in tight connection with the Galactic Bulge.
All the collected pieces of evidence point towards the possibility that Terzan 5 could be the
remnant of one of the pristine fragments that contributed to form the Bulge itself,
according to a scenario that has been theoretically proposed by several authors,
such as \cite{immeli} or \cite{elmegreen08}.

\section{Terzan 5 Metallicity Distribution}

\begin{figure}[!htp]
\resizebox{\hsize}{!}{\includegraphics[clip=true]{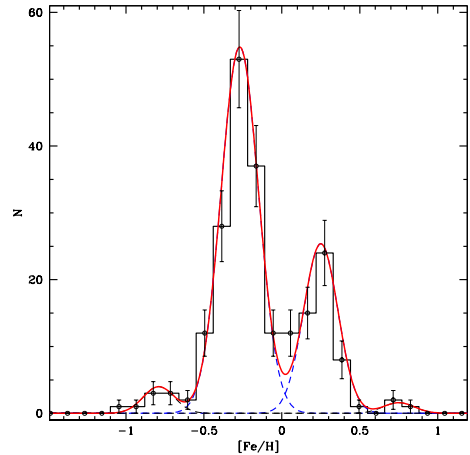}}
\caption{
\footnotesize
Metallicity distribution for a sample of 215 likely members giants of Terzan 5. The
distribution spans a range of metallicity larger than 1.5 dex and is clearly multimodal,
with the presence of possibly four peaks.
}
\label{distrib}
\end{figure}

Within this exciting scenario, and in order to better constrain the formation and evolution of
this system, we collected additional medium- and high- resolution spectra with DEIMOS@Keck and
FLAMES@VLT (with resolutions ranging from R$\sim7000$ for the spectra taken with DEIMOS, 
to R$\sim16000$ for those obtained with FLAMES) 
for a sample of more than 1600 stars located in the field of view of Terzan 5.
For a sub-sample of 215 stars, which are likely members of Terzan 5 according
to their radial velocities and locations within the tidal radius, we measured iron abundances.
For the 158 targets observed with FLAMES, we employed
the equivalent width method on a list of 12 FeI lines falling within the spectral 
range covered by the HR21 grating and derived the iron abundances by using the package GALA \citep{gala}. 
For 24 targets observed with DEIMOS, due to the lower resolution and the high
degree of line blanketing, the iron abundances have been measured by 
comparing the observed spectra with a grid of synthetic spectra, according to the
procedure described by \citet{m12_2419}. The remaining 33 stars are those analyzed
by \cite{origlia} in their work.
The resulting preliminary metallicity distribution is shown in Fig.\ref{distrib}

The distribution is extremely peculiar. First of all
it covers a huge range of metallicities, from [Fe/H]$\simeq-1$ dex, to
[Fe/H]$\simeq+0.9$ dex. The mean value of the distribution is [Fe/H]$\simeq-0.19$ dex
and its dispersion ($\sigma=0.31$) dex is much larger than the typical uncertainties
on the measured abundances. Second, it is clearly multi-modal, with the possible
presence of four different peaks.
The two main peaks are located at [Fe/H]$\simeq-0.3$ dex and [Fe/H]$\simeq+0.3$ dex and therefore
they well correspond to the two main populations already identified by \cite{origlia}.
The most metal-poor component, numerically corresponding to a few percent of the entire sample
($\sim$3\%), is located at about [Fe/H]$\simeq-0.8$ dex.
Recently, the analysis of near infrared high-resolution spectra (\citealt{origlia97}) 
acquired with NIRSPEC@Keck on three stars belonging to this metal-poorer 
component solidly confirmed its presence (\citealt{o13}). Its measured average iron abundance is [Fe/H]$=-0.79$ dex, 
thus enlarging the spread in metallicity covered by Terzan 5 up to $\Delta$[Fe/H]$>1$ dex. 
These authors have also been able to measure the $\alpha$-elements abundances for the three stars,
finding an average $\alpha$-enhancement of $[\alpha$/Fe]$=+0.36$ dex. This value is in very good agreement
with that of the main metal-poor component.
Finally the most metal-rich peak,
located at [Fe/H]$\simeq+0.7$ dex, is composed of only 3 stars ($\sim1.5\%$ of the sample)
and therefore we need a deeper investigation (Massari et al. submitted).

\section{Discussion}

There is only another GC-like system in the Galaxy that shows properties (in terms of
metallicity distribution) similar to those of Terzan 5: $\omega$ Centauri (see \citealt{omega}, 
\citealt{origlia03}, \citealt{sollima04, sollima07}, \citealt{bellini10}, \citealt{jp10},
\citealt{pancino2000}, \citealt{ferraro04, ferraro06}).
However, $\omega$ Centauri, which is now supposed to be the remnant of an accreted dwarf galaxy
(\citealt{jp10}), is located in a metallicity regime that is
much metal-poorer compared to that of Terzan 5. Moreover, also the $\alpha$-element abundance pattern
observed in Terzan 5 is completely different from that typically observed in
dwarf spheroidal and dwarf elliptical galaxies. Therefore, the extra-galactic origin of Terzan 5
is extremely unlikely.

Instead, the multimodality observed in the metallicity distribution of Terzan 5 well matches that observed for
Bulge stars in recent studies, such as \cite{ness} or \cite{bensby}, thus tightening
the already strong link between Terzan 5 and the Bulge suggested by the $\alpha$-elements abundances.
Therefore we conclude that Terzan 5 could be interpreted as the remnant of one of the pristine
fragments that contributed the form the Galactic Bulge.
Crucial information to possibly confirm this scenario will come from the kinematical properties of Terzan 5 
and from the search of similar systems among other GC-like systems in the Bulge.

\begin{acknowledgements}
This research is part of the project COSMIC-LAB 
(web site: www.cosmic-lab.eu) funded by the European Research 
Council (under contract ERC-2010-AdG-267675).
\end{acknowledgements}

\bibliographystyle{aa}

\end{document}